\documentclass[12pt]{article}

\usepackage{epsfig}

\newcommand{\ed}{\end{document}}
\newcommand{\be}{\begin{equation}}
\newcommand{\ee}{\end{equation}}
\newcommand{\ba}{\begin{eqnarray}}
\newcommand{\ea}{\end{eqnarray}}
\newcommand{\baz}{\begin{eqnarray*}}
\newcommand{\eaz}{\end{eqnarray*}}
\newcommand{\bb}{\begin{thebibliography}}
\newcommand{\eb}{\end{thebibliography}}

\begin{document}
\sloppy
\thispagestyle{empty}
\hfill {FTUV-01-0405;  IFIC/01-20}

\vspace{2cm}

\mbox{}

\begin{center}
{\large\bf   Evidence for the flavor singlet  axial anomaly related effects in 
$\phi$  meson electromagnetic production at large momentum
transfers}

\vskip 0.5cm

 N.I. Kochelev$^{1,2,3}$  and  V. Vento$^1$\\
\vspace{5mm}
{\small\it
$^1$ Departament de F\'{\i}sica Te\`orica and Institut de F\'{\i}sica 
Corpuscular,\\
Universitat de Val\`encia-CSIC, 
E-46100 Burjassot (Valencia), Spain \\
$^2$ BLTP,
Joint Institute for Nuclear Research,\\
Dubna, Moscow region, 141980 Russia\\
$^3$ Institute of Physics and Technology, Almaty, 480082, Kazakhstan}

\vspace*{1cm}

\end{center}
\begin{abstract}
The gluonic contributions to the conventional PCAC formulas 
due to flavor singlet axial anomaly have been instrumental in explaining
the mass of the $\eta^{\prime}$ and providing a plausible explanation 
for solving the spin crisis. We show that they also play an important
role in the description of photo- and electroproduction of
vector mesons at low energy and high momentum transfers. We calculate 
the  contributions of this type to $\phi$ meson electromagnetic production 
in a model, which contains also a soft pomeron, and
find agreement with recent CLAS data.
\end{abstract}

\vskip 1.0cm
\leftline{Pacs: 12.40.Nn, 13.60.Le,13.88.+e}
\leftline{Keywords: axial anomaly, vector mesons, gluon, large momentum 
transfer.}
\vspace{0.5cm}

{\tt
\leftline {kochelev@thsun1.jinr.ru}
\leftline{vicente.vento@uv.es}
}

\newpage

\section{Introduction}

The CLAS data on $\phi$ meson photo- and
electroproduction \cite{CLAS1, CLAS2} at large momentum
transfers are widely under discussion
\cite{laget,titov,zhao}.  The main interest lies in
unveiling the structure of the nucleon, in particular its
strange quark content\cite{anselmino,dorokhov}. The
relation between the structure of the nucleon, the so
called proton spin problem, and 
the gluonic contributions related to the flavor singlet axial 
anomaly, hereafter abreviated by FSAA, have been understood for some 
time \cite{anselmino}. The latter
also furnishes an explanation for OZI rule violations,  
describing why decays of heavy resonances that involve
disconnected quark graphs, i.e. graphs that can only be
connected via gluon lines, are suppressed. Thus it is to be
expected that the FSAA mechanism be manifest in those
channels of vector meson electromagnetic production where
gluonic degrees of freedom are important. 

The FSSA arises as a consequence of the
complicated vacuum structure of QCD. The breaking of the
$U_A(1) $ symmetry in the theory requires a complex
mechanism, associated with the periodicity of the QCD
potential as a function of the topological charge, producing
a massless pole in the  correlator of two anomalous
currents  \cite{diakonov}. The pole is not associated to
any physical particle because the currents are gauge
dependent. It was first introduced by Veneziano
\cite{veneziano1} to explain the large $\eta^\prime$ mass in
the framework developed by Witten \cite{witten}. 

The role of the FSAA in understanding the structure
of the nucleon has been pointed out by Shore and Veneziano 
\cite{veneziano2}. We consider its contribution  to $\phi$
meson elastic electromagnetic production off the nucleon,
which is a suitable scenario to investigate the OZI rule.
In $\phi$ meson production the contribution of the valence
quark exchange is much smaller than in $\rho$  and $\omega$
meson production, therefore this reaction gives the
opportunity to investigate the role of the gluonic
exchanges in vector meson production.

\section{ The dynamics associated with the FSAA mechanism}

The $U_A(1)$ symmetry of QCD is anomalous. The anomaly is
reflected in the divergence of the current by the appearence
of a term proportional to the topological charge density
operator

\begin{equation}
Q(x)=\frac{\alpha_s}{8\pi}G_{\mu\nu}^a(x)\widetilde{G}_{\mu\nu}^a(x).
\label{topol}
\end{equation}
The axial singlet Ward identities are anomalous
and the mechanism which provides the large mass to
the $\eta^{\prime}$ is the non vanishing of the zero momentum
correlator of two topological charges. As a consequence the
anomalous gluonic current

\begin{equation} 
K_{\mu}=\frac{\alpha_s}{4\pi}\epsilon_{\mu\nu\alpha\beta}A_\nu^a
(\partial_\alpha
A_\beta^a+\frac{g_s}{3}f_{abc}A_\alpha^bA_\beta^c)
\label{current}
\end{equation}
has a massless pole 

\begin{equation}
{i\int d^4x
e^{ikx}<0|TK_{\mu}(x)K_{\nu}(0)|0>_{k\rightarrow0}}\rightarrow
\frac{g_{\mu\nu}}{k^2}\lambda^4 
\label{correlator}
\end{equation} 
which is related to the topological susceptibility of the
QCD vacuum

\begin{equation}
\chi(0)=-\lambda^4=i\int d^4x<0|TQ(x)Q(0)|0>.
\label{correlator2}
\end{equation}

The matrix elements between the $Q$-state and the pseudoscalar 
isosinglet quark states   \cite{diakonov}
\begin{equation}
\phi_5^i=\bar q_i\gamma_5 q_i,
\label{states}
\end{equation}
where $i=u,d,s$, are non zero. The $Q$-operator allows
transitions  between states with different quark flavors
and provides the mechanism  for the observed large OZI violation 
in $\eta-\eta^\prime$ system. 

The diagonalization of the propagator matrix for the 
operator $Q$ and  quark nonet states $\pi^0,\eta^8$ and $\eta^\prime$
leads to the following gauge invariant propagator
\begin{equation}
<GG>=-A,
\label{ghost}
\end{equation}
where $G$ is the linear combination of $Q$ and $\phi_5^i$ operators.
For the physical mesons the propagator is given by 
\begin{equation}
<\eta^\alpha\eta^\beta>=\frac{-\delta^{\alpha\beta}}{k^2-m_{\eta^\alpha}^2},
\label{eta}
\end{equation}
where $\eta^\alpha=\eta^\prime, \eta, \pi^0$\cite{shore}.

$A$ in (\ref{ghost}) is determined from
the topological susceptibility by the equation
\begin{equation}
\chi(0)=-A(1-A\sum_q\frac{1}{m_q<\bar qq>})^{-1}.
\label{hi}
\end{equation}
We approximate $A$ by the susceptibility of the Yang Mills
theory, $\chi(0)\vert_{YM}$, and take for the latter
the  value given by the Veneziano-Witten formula
formula\footnote{The lattice calculations of the
topological
susceptibility   \cite{giacomo} get
$\chi(0)=-(175\pm 5$ MeV)$^4$, confirming this value.}  
\cite{witten,veneziano1}  
\begin{equation}
\chi(0)\vert_{YM}
=-\frac{f_\pi^2}{6}
(m_{\eta^\prime}^2+m_\eta^2-2m_K^2) \approx (180 MeV)^4.
\label{top}
\end{equation}

We should stress that the $G$-propagator (\ref{ghost})
does not depend on the transfer momentum $k^2$, therefore
the effective interaction induced by the $G$- exchange is
a {\it point-like} interaction. Below we shall show that
this property is responsible for the large $G$- pole 
contribution to the electromagnetic $\phi$ meson production
at large momentum transfers.

The contribution of the $G$-pole to the physical
amplitudes leads to modifications on the predictions by the
OZI rule. An example where this mechanism is at work, of
interest here, is the generalized $U(1)$ Golberger-Treiman
relation for isosinglet axial-vector nucleon form factor at
zero momentum transfer \cite{veneziano1} 

\begin{equation}
2M_NG_A^0=Fg_{\eta^\prime NN}+2N_fAg_{GNN}=F_{\eta_0}g_{\eta_{0NN}},
\label{GT}
\end{equation}
where $F\approx\sqrt{2N_f}f_\pi$, $f_\pi=93$ MeV. 
The $G$-nucleon coupling constant $g_{GNN}$ has been
defined by the interaction
\begin{equation}
L_{QNN}=ig_{GNN}G\bar N\gamma_5 N.
\label{GNN}
\end{equation}

One can interpret the formula (\ref{GT}) as establishing a
relation between the coupling constant of the physical $\eta^\prime$ meson 
and that of the nonphysical $\eta_0$ meson in OZIQCD.
The $G$-pole contribution in (\ref{GT}) just describes the OZI
violating piece of the coupling due to intermediate gluonic states
(see Fig.1).
\begin{figure}[htb]
\centering
\epsfig{file=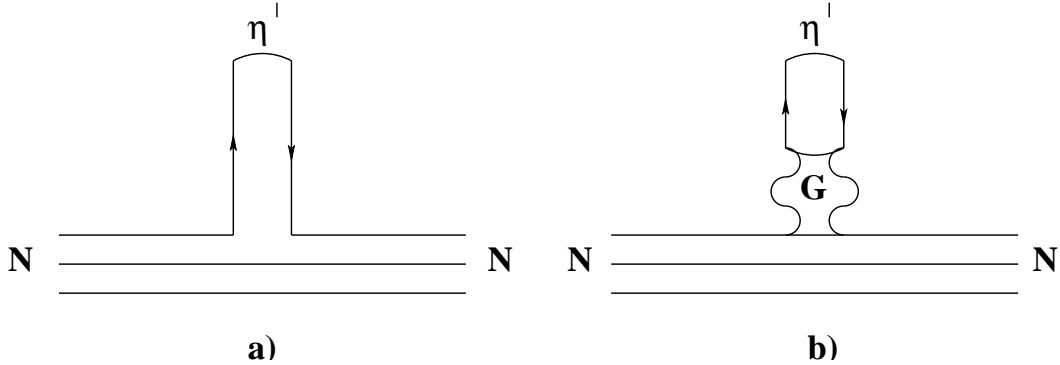,width=14cm}
\vskip 0.5cm
\caption{\it The contributions to the $\eta^\prime-NN$ interaction
a) due to OZI conserving piece and b) the gluonic $G$-pole contribution. }
\end{figure}

The contribution of the $G$-pole leads to a significant
reduction of the flavor singlet  axial charge from naive
OZI expectation of $G_A^0\approx G_A^8\approx 0.6$ and
provides an explanation of the famous {\it spin crisis}
(see reviews \cite{anselmino,dorokhov}). The contribution
should be negative and its effect  in (\ref{GT})  is to 
make $G_A^0$ reach its experimental value of $\approx 0.3$.
This allows us to estimate the value of $G$-nucleon
coupling constant, for $N_f =3$, as 

\begin{equation}  
g_{GNN}\approx-\frac{0.3M_N}{N_fA}\approx-89.35 GeV^{-3}.
\label{gGNN} 
\end{equation}

The $G$-pole contribution to the $\eta^\prime(\eta)\rightarrow\gamma\gamma$ 
decay has been discussed by Shore \cite{shore}. Defining the
interaction vertex as

\begin{equation}
<\gamma\gamma|\eta^\prime>=-ig_{\eta^\prime\gamma\gamma}   
\epsilon_{\mu\nu\alpha\beta}k_1^\mu k_2^\nu\epsilon^
\alpha(k_1)\epsilon^\beta(k_2),
\label{vertex}
\end{equation}
a modified formula  for the  effective coupling of the $\eta^\prime$ meson 
with photons has been obtained,
\begin{equation}
Fg_{\eta^\prime\gamma\gamma}+2N_fAg_{G\gamma\gamma}=
\frac{4}{\pi}\alpha_{em}.
\label{Ggg}
\end{equation}
One can understand this relation in a similar way as for the
effective $\eta^\prime$-nucleon coupling. There are two 
ways for the coupling to take place: i) an OZI preserving
one; ii) another which incorporates OZI violating terms
determined via the $G$-pole
(Fig.2).
\begin{figure}[htb]
\centering
\epsfig{file=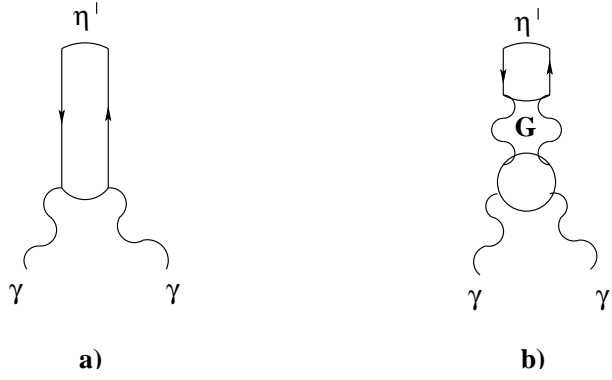,width=8cm}
\vskip 0.5cm
\caption{\it The contributions to the $\eta^\prime-\gamma\gamma$ decay
a) OZI conserving part and b) $G$-pole contribution. }
\end{figure}

Let us consider the $G$-pole contribution to the $\eta^\prime\gamma\phi$
coupling
\footnote{
Similar considerations can be perform  for $\rho$ and 
$\omega$ mesons .}. The interaction of the $\phi$ meson with
the
s quarks is assumed to be photon-like
\begin{equation} 
L_{\phi ss}=C_\phi\bar s\gamma_\mu s\phi_\mu .
\label{phiss}
\end{equation}
Due to this vector meson-photon anology the generalization of (\ref{Ggg}) 
to the case of $\eta^\prime\rightarrow\gamma \phi$ 
decay is straightforward (Fig.3)
\begin{figure}[htb]
\centering
\epsfig{file=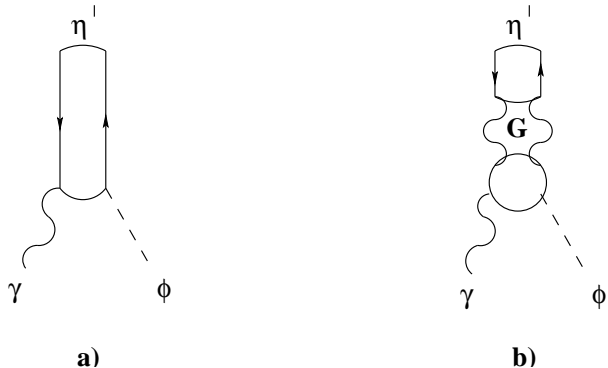,width=8cm}
\vskip 0.5cm
\caption{\it The contributions to the $\eta^\prime-\gamma\phi$ vertex
a) OZI conserving part and b) $G$-pole contribution. }
\end{figure}
\begin{equation}
Fg_{\eta^\prime\gamma\phi}+2N_fAg_{G\gamma\phi}=
-C_\phi\sqrt{\frac{\alpha_{em}}{\pi^3}}.
\label{GgPhi}
\end{equation}
The value of $g_{\eta^\prime\gamma\phi}$ can be
extracted from experimental width 
$\Gamma_{\phi\rightarrow\eta^\prime\gamma}=2.99\times 10^{-4}$ MeV
\cite{PDG}.
Using the calculated decay width
\begin{equation}
\Gamma_{\phi\rightarrow\eta^\prime\gamma}=
\frac{(M_\phi^2-M_{\eta^\prime}^2)^3}{96\pi M_\phi^3}
g_{\eta^\prime\gamma\phi}^2,
\label{width}
\end{equation}
we obtain
\begin{equation}
|g_{\eta^\prime\gamma\phi}|=0.233 GeV^{-1}.
\label{gegp}
\end{equation}

Unfortunately the $\phi$-meson coupling with the strange
quarks 
is not well known. In order to estimate it we use three
scenarios. The naive quark model (NQM)  calculation provides us
with

\begin{equation}
C_\phi=g_{\phi KK}\approx\frac{g_{K^+K^-}+g_{K_LK_S}}{2}=4.53,
\label{NQM}
\end{equation}
where the values of the $\phi-KK$ coupling from 
\cite{achasov} was used.
The vector meson dominance (VMD) model at the quark level gives
\begin{equation}
C_\phi=\frac{g_{\phi}}{3}\approx 4.37,
\label{VDM}
\end{equation}
where $g_{\phi}=13.1$ is determinated from electronic width of $\phi$ meson.
The last estimate comes from the Nambu-Jona-Lasinio (NJL) model \cite{volkov}
\begin{equation}
C_\phi=5.33.
\label{NJL}
\end{equation}
Our procedure does not determine the sign of the some of 
coupling constants.
However in the large $N_c$ limit of the theory, the sign of the first term 
in (\ref{GgPhi}), should be the same as that of the OZIQCD term in the right 
hand side of the equation. Therefore we can write,
\begin{equation}
|g_{G\gamma\phi}|=
  \frac{|Fg_{\eta^\prime\gamma\phi}|-|C_\phi\sqrt{\alpha_{em}/\pi^3}|}
{2N_fA}
\label{couplG}
\end{equation}
which gives $|g_{G\gamma\phi}^{NQM}|=2.88$ GeV$^{-4}$,
$|g_{G\gamma\phi}^{VDM}|=3.08$ 
GeV$^{-4}$ and $|g_{G\gamma\phi}^{NJL}|=1.91$ GeV$^{-4}$.

\section{$G$-pole  contribution to the $\phi$ meson photo-
and electroproduction}
It is very well known that at large energy and small 
momentum transfer the main contribution to the elastic electromagnetic
production of the vector mesons comes from the pomeron exchange
(see \cite{DL1} and references therein).
At low energies the pomeron still gives the main
contribution to the
$\phi$ production cross section since OZIQCD is valid for reactions at 
small momentum transfers \cite{laget,titov,zhao} 
(Fig.4a).
\begin{figure}[htb]
\centering
\epsfig{file=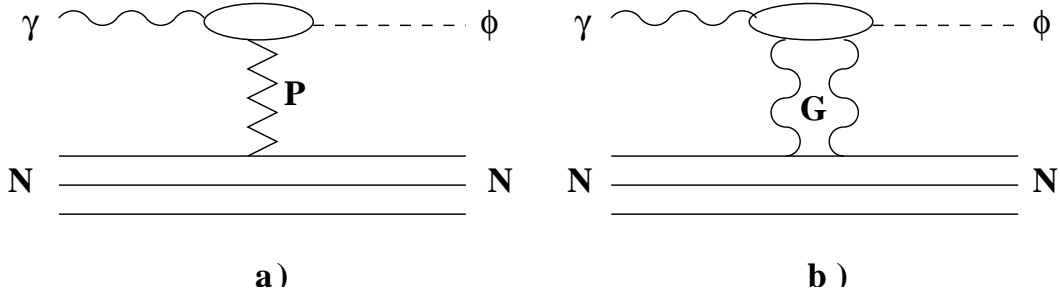,width=14cm}
\vskip 0.5cm
\caption{\it The contributions of the a) pomeron
b) $G$-pole exchange to the electromagnetic $\phi$-meson production. }
\end{figure}
We use the Donnachie-Landshoff (DL) model \cite{DL2} to
describe the soft 
pomeron contribution to the elastic $\phi$ meson production
differential cross section.
This model is based on  two gluon exchange contributions to
 quark-quark scattering \cite{LN}.

In the DL model the contribution of the pomeron to the
elastic cross section is given by 
\begin{equation}
\frac{d\sigma_P}{dt} = \frac{81 m_\phi^3 \beta_0^2\beta_s^2 
\Gamma_{e^+e^-}^\phi}
{\pi \alpha_{\rm em}}
\frac{F(t)^2}{(2\mu_0^2 + Q^2 + m_\phi^2 - t)^2(Q^2 + m_\phi^2 - t)^2}
 \left( \frac{S}{S_0} \right)^{2 \alpha_P(t) - 2},
\label{cross3}
\end{equation}
where 
\begin{equation}  
F(t)=\frac{4M_N^2-2.8t}{(4M_N^2-t)(1-t/0.7)^2}
\label{form1}
\end{equation}
is the electromagnetic nucleon form factor,
and $\beta_0=2$ GeV$^{-1}$, $\beta_s=1.5$ GeV$^{-1}$, $\mu_0=1.2$ GeV, 
$S_0=1/\alpha_P^\prime$ and pomeron trajectory is
$\alpha_P(t)=\alpha_P(0)+\alpha_P^\prime t$, with 
$\alpha_P(0)=1.08$ and 
$\alpha_P^\prime=0.25$ GeV$^{-1} $, $S=W^2$. 
The contribution of the DL pomeron to the cross section for the
$\phi$ meson photo- and electroproduction for CLAS kinematics
is presented in Figs. 5 and 6 by dashed lines.
\begin{figure}[htb]
\centering
\epsfig{file=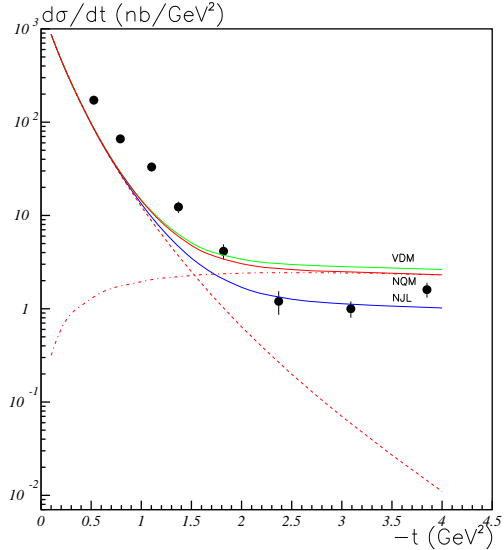,width=8cm}
\vskip 0.5cm
\caption{\it The contributions of the  pomeron (dashed line) and
$G$-pole exchange (dot-dashed line) with NQM coupling to the  
$\phi$-meson photoproduction
at $W=2.76 $ GeV. Solid lines are the total contribution 
for the different $\phi-s$ quark couplings. The data are
from the CLAS 
collaboration \cite{CLAS1}.}
\end{figure}
\begin{figure}[htb]
\centering
\epsfig{file=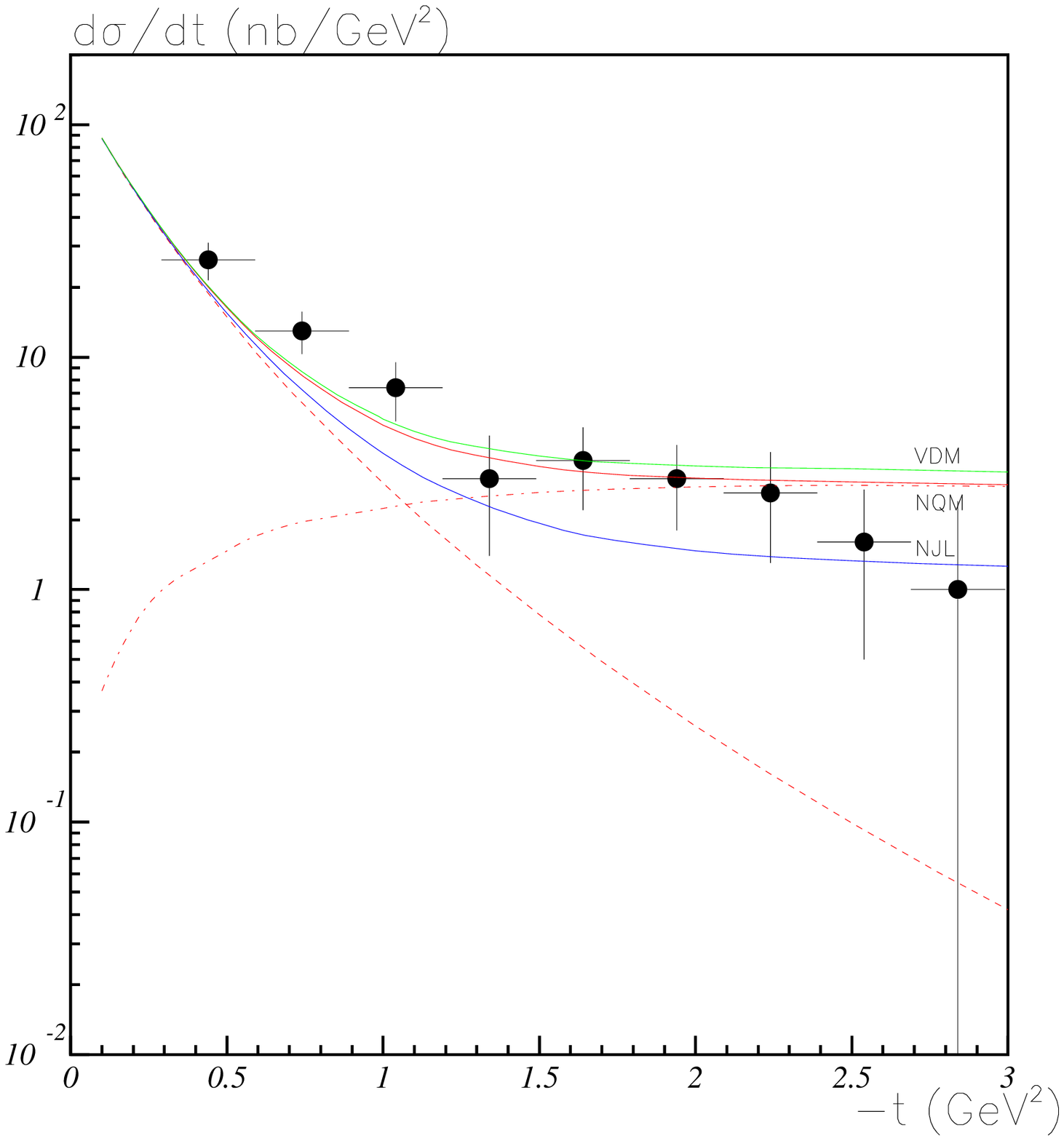,width=8cm}
\vskip 0.5cm
\caption{\it The contribution of the  pomeron and
 $G$-pole exchange to the  $\phi$-meson electroproduction for 
$<Q^2>=1.45$ GeV$^2$ and $W=2.3$ GeV . The data are from the CLAS 
collaboration \cite{CLAS2}. The notation is the same as in Fig.5. }
\end{figure}

For small $-t\leq 1$ GeV$^2$ the pomeron  contribution 
describes the data rather
well, but for large $-t\geq 1$ GeV$^2$, the deviation from the
experiment is large. Furthermore, the electroproduction data show some 
evidence for a dip near $-t\approx 1$ GeV$^2$, which is impossible
to obtain within the DL model. There are some attempts  to describe large
$t$ CLAS photoproduction data by  the 
$u-$channel contributions \cite{laget,zhao}. But in this case one should expect
a large suppression of this contributions in $\phi$-meson 
 electroproduction due to the strong  $Q^2$ dependence of the 
form factors. This suppresion is not seen in the CLAS data (Fig.6).

The main idea behind our approach to the large $-t$ behavior
of the cross section is  that the $G$-pole exchange mechanism
will soften the decrease of the cross section from the
pomeron. The reason is very simple, the $G$ propagator (\ref{ghost}) does 
not have any $t$ dependence  \cite{anselmino}. From the
analysis of the nucleon spin dependent structure function $g_1(x,Q^2)$
we know that the $G$-pole contribution, which can be considered
as the gluon  contribution to structure function, has only
weak (logarithmic) $Q^2$ dependence \cite{anselmino}.
Therefore the $G$-pole exchange should give the flattening
at large $-t$ of the cross section  in both photo- and electroproduction. 
The absolute value of the $G$-pole contribution (Fig.4b)  strongly
depends on its couplings with nucleon, photon and $\phi$-meson
and is given by 
\footnote{We neglect the possible weak  $Q^2$ dependence of the 
$G$-pole contribution to the matrix element of the reaction.}
\begin{equation}
\frac{d\sigma_{G}}{dt} = -\frac{A^2g_{G\gamma\phi}^2g_{GNN}^2
t(t-M_\phi^2)^2}{64\pi((S+Q^2+M_N^2)^2-4SM_N^2)}F_1(t)^2,
\label{gcross}
\end{equation}
where $F_1(t)$ is the flavor singlet axial form factor of the nucleon
which we take of the form (see \cite{kochelev})
\begin{equation}
F_1(t)=\frac{1}{(1-t/M_{f_1}^2)^2}.
\label{form2}
\end{equation}
In (\ref{form2}) $M_{f_1}=1.285$ GeV  is the mass of the flavor singlet 
$f_1$  meson. 

It should be noted that the $G$-pole exchange  induces a nucleon
spin-flip, which  produces a factor $t$ in
Eq.(\ref{gcross}), and leads to an  additional  enhancement
of the $G$ contribution at large $-t$, as compared with
pomeron contribution, which is nonspin-flip
(\ref{cross3}).  Moreover the energy dependence of the
$G$ contribution corresponds to a fix pole with zero
Regge slope.  Therefore the  large $t$ Regge suppression 
to vector meson production
given by $({S}/{M_N^2})^{2\alpha_R^\prime t}$ with
slope  $\alpha_R^\prime\approx 0.9$ GeV$^{-2}$ for the  
usual Regge trajectories, e.g. 
$\pi^0$ and $\eta $, is
absent for the $G$-pole exchange.

The result of the calculation of the $G$ contribution
with the NQM  $\phi-$strange quark  coupling  is presented
in Figs. 5 and 6 by the dot-dashed line. In them  the sum of the
pomeron and $G$ contributions
within all models for the couplings are shown by the solid lines.
It is evident that $G$ contribution determines the behavior
of the cross section at large $-t$. The agreement with the data can 
be, of course, improved by  the inclusion of another meson exchanges,
for example $\eta$ , $\pi^0$ and $f_1$ , but already our simple
model based only on pomeron and $G$-pole contributions reproduces 
the main features of the CLAS data.

\section{Conclusion}
We have shown that  gluonic degrees of freedom 
play a very important  role
in the electromagnetic $\phi$-meson production at small
energy and large $-t$.
At small $-t$ the cross-section is described rather well by
the
Donnachie-Landshoff soft pomeron model.
At large $-t$, both in the photo- and electroproduction cross
sections,
an  extremely interesting phenomena related to the
complex topological structure of QCD vacuum, takes place.
We have shown that at large momentum transfer, the point-like
interaction induced by the  axial anomaly
gives the dominant contribution.   
It should be mentioned that a similar phenomena should appear also 
in $\rho$, $\omega$ and $J/\Psi$ production at low energy and large
momentum transfer, although it might be masked by other
mechanisms \cite{kv}. We expect, though, that the sum of pomeron and 
$G$-pole contributions  to $J/\Psi$  photoproduction might be
dominant near 
threshold, because here the 
minimal momentum transfer is large, $-t_{min}=2.2$ GeV$^2$.
This mechanism might explain the unusual
flat  energy dependence of the $J/\Psi$ cross section observed at    
Cornell \cite{cornell} for $E_{\gamma} \leq 12 GeV$.

The $G$ exchange is an unnatural parity exchange. Therefore it can be
separated from the usual hard gluonic exchange
at large $-t$
\cite{laget2} by looking at the angular distribution of the
$\phi$ decay .
This procedure was suggested in \cite{kochelev2} to disantangle
the anomalous unnatural parity exchange $f_1$ and the
natural parity hard
pomeron 
contribution in vector meson photoproduction at large energies. 
In this connection we should point that CLAS photoproduction 
data \cite{CLAS1} show a big change of the $K$-mesons 
angular distribution at large $-t$, which might be a signal
for a 
large $G$-pole exchange contribution.

From our point of view vector meson  
electroproduction at large momentum 
transfer opens the new opportunity to investigate the complex structure
of the QCD vacuum.

\section*{Acknowledgements}
We are grateful to  A.E.Dorokhov, S.B. Gerasimov, N.N.Achasov 
and V.L.Yudichev for 
useful discussion. We thank R. Schumacher for providing us
with the CLAS data.
One of us (N.I.K) is  grateful to
the University of Valencia for the warm hospitality. 
This work was partially  supported by DGICYT PB97-1227, RFBR-01-02-16431
and INTAS-2000-366 grants.


\begin{thebibliography}{99}

\bibitem{CLAS1} CLAS Collaboration, E.Anciant  et al.,
 Phys. Rev. Lett. {\bf 85} (2000) 4682.
\bibitem{CLAS2} CLAS Collaboration, K.Lukashin et al.,
hep-ex/0101030.
\bibitem{laget} J.-M.Laget,   Phys. Lett. {\bf B489} (2000) 313.
\bibitem{titov} Y.Oh, A.I.Titov and T.-S.H.Lee, nucl-th/0004055;\\
Y.Oh, A.I.Titov, S.N.Yang, T.Morii,
 Nucl. Phys. 
 {\bf A684} (2001) 354.
\bibitem{zhao} Q.Zhao, B.Saghai and J.S.Al-Khalili, nucl-th/0102025.
\bibitem{anselmino} M.Anselmino, A.Efremov and E.Leader,
Phys. Rep. {\bf 261} (1995) 1.
\bibitem{dorokhov} A.E.Dorokhov,N.I.Kochelev and A.Yu.Zubov,
Int. J. of Mod.Phys. {\bf A8} (1993) 603.
\bibitem{diakonov} D.I.Dyakonov and M.I.Eides, Sov. Phys. JETP. {\bf 54} 
(1981) 2.
\bibitem{veneziano1} G.Veneziano, Nucl. Phys. {\bf B159} (1979) 213.
\bibitem{witten} E.Witten, Nucl. Phys.{\bf B156} (1979) 269.
\bibitem{veneziano2} G.Veneziano, Mod. Phys. Lett. {\bf A4} (1989) 1605;\\
G.M.Shore and G.Veneziano, Nucl. Phys.{\bf B381} (1992) 23.
\bibitem{shore}
G.M.Shore, Nucl. Phys. {\bf B569} (2000) 107
and hep-ph/9908273.
\bibitem{giacomo}
B.Alles,  M.D'Elia and A. Di Giacomo, Nucl.Phys. {\bf B494} (1997) 281.
\bibitem{achasov} M.N.Achasov et. al., hep-ex/0009036. 
\bibitem{volkov} M.K.Volkov, Sov.J. Part.Nucl. {\bf 17} (1986) 282;\\
D.Ebert, M.K.Volkov and V.L.Yudichev, J. Phys. {\bf G25} (1999) 2025.
\bibitem{PDG}
\mbox{Particle Data Group,} C.~Caso {\em et~al.\/},
  Eur. Phys. J. C {\bf 3}, 1 (1998).
\bibitem{DL1} A.Donnachie and P.V.Landshoff, Phys. Lett. {\bf B478}
(2000) 146.
\bibitem{DL2} A.Donnachie and P.V.Landshoff, Phys. Lett. {\bf B185}
(1987) 403
\bibitem{LN} F.E.Low, Phys.Rev. {\bf D12} (1975) 163;\\
S.Nussinov, Phys.Rev.Lett.{\bf 34} (1975) 1275;\\
P.V.Landshoff and O.Nachtmann, Z. Phys. {\bf C35} (1987) 405.
\bibitem{kochelev}
N.~I. Kochelev, D.-P. Min, Y.~Oh, V.~Vento, and A.~V. Vinnikov,
  Phys. Rev. D {\bf 61}, (2000) 094008.
\bibitem{kv} N.I. Kochelev and V. Vento, in preparation.
\bibitem{laget2} J.-M. Laget and R.Mendez-Galain,
Nucl. Phys. {\bf A581} (1995) 397.
\bibitem{kochelev2}
Y.~Oh, N.~I. Kochelev, D.-P. Min, V.~Vento, and A.~V. Vinnikov,
  Phys. Rev. D {\bf 62}, (2000) 017504.
\bibitem{cornell} B. Gittelman et al. Phys. Rev. Lett. {\bf 35} (1975) 1616.
\end{thebibliography}
\end{document}